# Fast detection of water nanopockets underneath wet-transferred graphene

Michele Magnozzi[a], Niloofar Haghighian[a], Vaidotas Miseikis[b], Ornella Cavalleri[c], Camilla Coletti[b, d], Francesco Bisio[e, *], Maurizio Canepa[a]

[a] OptMatLab, Dipartimento di Fisica, Università di Genova, via Dodecaneso 33, 16146 Genova, Italy

[b] CNI@NEST, Istituto Italiano di Tecnologia, Piazza S. Silvestro 12, 56127 Pisa, Italy

[c] Dipartimento di Fisica, Università di Genova, via Dodecaneso 33, 16146 Genova, Italy

[d] Graphene Labs, Istituto Italiano di Tecnologia, via Morego 30, 16163 Genova, Italy

[e] CNR-SPIN, C.so Perrone 24, 16152 Genova, Italy

## Abstract

We report an investigation of the graphene/substrate interface morphology in large-area polycrystalline graphene grown by chemical-vapour deposition and wet-transferred onto Si wafers. We combined spectroscopic ellipsometry, X-ray photoelectron spectroscopy and atomic-force microscopy in order to yield morphological and chemical information about the system. The data showed that wet-transferred samples may randomly exhibit nanosized relief patterns indicative of small water nanopockets trapped between graphene and the underlying substrate. These pockets affect the adhesion of graphene to the substrate, but can be efficiently removed upon a mild annealing in high vacuum. We show that ellipsometry is capable of successfully and reliably detecting, *via* multilayer dielectric modelling, both the presence of such a spurious intercalation layer and its removal. The fast, broadly applicable and non-invasive character of this technique can therefore promote its application for quickly and reliably assessing the degree of adhesion of graphene transferred onto target substrates, either for *ex-post* evaluation or in-line process monitoring.

## 1. Introduction

Large-area polycrystalline single-layer graphene (SLG) can be fabricated with great efficiency by chemical-vapour deposition (CVD) onto catalyst-metal substrates for most practical applications [1-4]. The SLG has to be softly detached from the catalyst metal in order to be transferred onto the target substrates, a delicate process that may involve spin-coating with polymers and the interaction of the system with both etchants and solvents. The functional response of the final system critically depend on the degree of interaction with the target substrate, the integrity of SLG, the presence of adventitious material either above SLG or between the substrate and SLG, to name a few [5,6]. The presence of intercalated material is of particular concern whenever wet transfers are involved, as either water or other solvents may remain trapped between graphene and the target substrate [7,8]. For technological purposes, beside the routine techniques for evaluating graphene's quality (*e.g.* Raman spectroscopy, scanning electron microscopy etc. [9]), it could be helpful to have a method available for assessing the presence of intercalated layers which is fast and reliable, non-invasive, with sensitivity down to the nanometer and sub-nm thickness and applicable in a number of different sample environments. In this work, we report on the application of spectroscopic ellipsometry (SE) for evaluating the morphology of CVD-grown graphene transferred onto a conventional $SiO_2$/Si wafer by wet methods. SE is a fast, reliable, nondestructive optical technique, applicable in air, vacuum and transparent liquids, including chemically-aggressive environments [10]. It is broadly exploited for the investigation of optical properties of materials and for *in-situ*, real-time monitoring of growth and conditioning processes in the scientific and technological field [11]. SE is extremely sensitive to even tiny variations in the morphology or optical response of surfaces, films and multilayers, that can be physically interpreted by means of suitable dielectric models [12-14]. Here we exploit these capabilities to quantitatively assess the presence of few-nm-thick water nanopockets trapped between transferred SLG sheets and the $SiO_2$ surface. The SE interpretation is based on an a dielectric-layer model including SLG, intercalation layer and substrate, whose outcome coherently agrees with corresponding atomic-force microscopy (AFM) and X-ray photoelectron spectroscopy (XPS) data. Altogether, the data show that mild annealing in vacuum effectively removes the nanopockets and improves the adhesion of SLG to the target substrate. Our work therefore promotes SE as a fast and reliable method for highlighting fine details of the graphene/substrate interfacial morphology and suggests this technique as a potential candidate for quickly assessing the final yield of SLG transfer procedures.

## 2. Results and discussion

### *2.1.    Experimental results*

Large-area continuous sheets of polycrystalline single-layer graphene ($\approx$10x10 $mm^2$) were synthesized by means of CVD on Cu foils, as described in detail in previous publications [15,16]. The samples were coated with a polymethylmethacrylate (PMMA) carrier membrane; the copper growth substrate was then etched using iron (III) chloride. The membrane was then rinsed in deionised water to remove etchant residues, and the target substrate (a commercial Si wafer with a nominal 290-nm-thick thermal-oxide layer) was used to pick up the membrane from the water, followed by the removal of PMMA in acetone. The experiments were performed on these systems as follows: the pristine samples were first characterized by SE, AFM and XPS.

The samples were then annealed under high-vacuum (HV) conditions ($p \approx 10^{-7}$ mbar) up to T=350 °C within a custom-made chamber designed for real-time SE measurements [17]. Following the annealing, the samples were again characterized by the same techniques.

SE data were collected by means of a J.A.Woollam M-2000 rotating compensator spectroscopic ellipsometer (spectral range 245-1700 nm, 0.73-5 eV). The SE data reported here have been measured *ex-situ*, in air at an incidence of 66° to match the angle allowed by the *in-situ*-SE apparatus. A typical SE spectrum can be acquired in a few seconds. The optical modelling was performed using the software WVASE32 provided with the instrument. XPS spectra were acquired by means of a PHI ESCA 5600 system, with monochromatized Al K$\alpha$ radiation. The AFM images were acquired by means of a Multimode/Nanoscope IV system, Digital Instruments Veeco, in tapping mode.

In Fig.1 (a) we report a representative AFM image (1x1 $mm^2$) of a pristine SLG sample. The image shows a peculiar morphology, unexpected for SLG laid onto a flat surface. Indeed, the height histogram of this image, reported in the inset of Fig. 1(a), shows a characteristic double-peaked structure, suggesting that the SLG/$SiO_2$ is morphologically a two-level system. The bottom level occupies around 35-40% of the surface area, and has a lower *rms* roughness with respect to the upper level (0.7 nm *vs* 1.4 nm). Overall, the *rms* roughness all over the image due to the SLG wrinkles, corrugation and two-level morphology reads 1.8 nm. The two levels are spaced in height by ~2 nm. The two-level morphology is not observed in all our samples. It randomly appears in a fraction of wet-transferred large-sheet samples, and is absent in specimens transferred by means of dry-transfer procedures.

The SE spectrum of the sample is reported in Fig. 1(c). The Ψ(*E*) and Δ(*E*) spectra (thick red and blue lines, respectively) are reported as a function of the photon energy *E*. The spectra are dominated by characteristic oscillations due to the thin-film interference arising from the optically-thick $SiO_2$ layer present on top of the Si substrate. The presence of SLG on top of the $SiO_2$/Si system leads to a general redshift of the Ψ and Δ oscillations with respect to the bare substrate (thin lines), and to a clear modification of the curve shape, specifically in the ultraviolet range (we notice that the redshift of the oscillations is readily interpreted as a thickness increase of the dielectric layer present on top of Si, coherent with expectations). Upon proper modelling, these spectra provide a wealth of information about the system as a whole, as shown below.

In Fig. 1(b) the energy regions corresponding to the O1s and C1s core levels in XPS are reported. The experimental data are reported as the open black markers, while the black line represents the best fit to the data. The XPS peaks have been decomposed into subcomponents related to different chemically-bound atomic states (coloured areas). The graphitic C1s peak (the dominant component of the C1s spectrum) was fitted as a Doniach-Sunjic singlet. The remaining C1s subcomponents, as well as the O1s subcomponents, were fitted as a superposition of Lorentian-Voigt components constrained to have, for each element, the same spectral width. A Shirley-type background was employed in all cases. The binding energy (BE) was referenced setting the large graphitic C1s component at 284.5 eV [18].

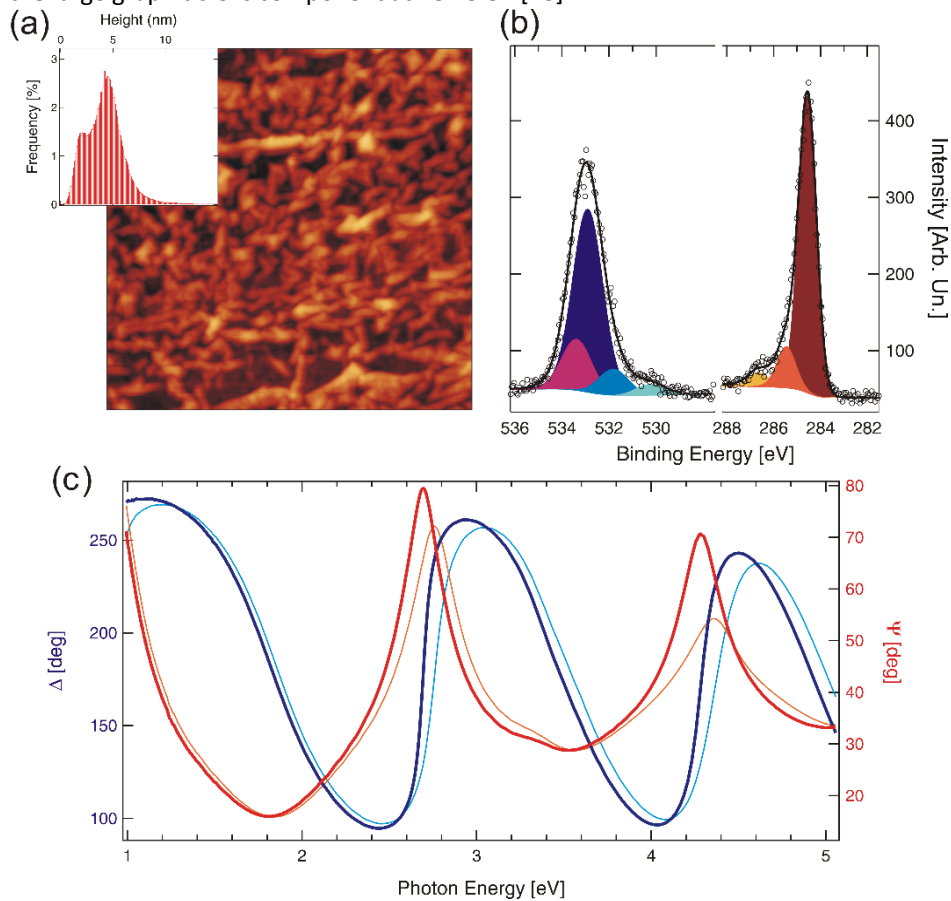

**Fig. 1.** Panel (a): AFM image of pristine SLG/$SiO_2$ exhibiting a two-level morphology (see the text for a detailed description of the system morphology, 1x1 mm$^2$). Inset of panel (a): corresponding height-

distribution histogram. Panel (b): high-resolution XPS spectra of the O1s (left) and C1s (right) spectral region of the pristine SLG/SiO$_2$ sample. The experimental data (markers), best fit (black line) and the fitting subcomponents (coloured areas) are shown. Panel (c): ellipsometry spectra Ψ (red line) and Δ (blue line) as a function of photon energy measured from the pristine SLG/SiO$_2$. The thin orange and light-blue lines are the Ψ(E) and Δ(E) spectra of the bare SiO$_2$/Si substrate. The incidence angle was 66°.

As previously mentioned, the carbon spectrum is dominated by one large graphitic subcomponent (brown), with two minor subcomponents, at 285.5 eV (orange) and 286.6 eV (dark yellow), respectively. The O1s spectrum shows a dominant component at 532.9 eV BE (dark blue, corresponding to oxygen bound in SiO$_2$) while several subcomponents appear around the major peak at 530.2 eV, 531.8 eV and 533.4 eV BE. The assignment of the subcomponents will be discussed in more detail in the following.

In Fig. 2 we report the AFM, SE and XPS measurements recorded *ex-situ* at room temperature following the sample annealing in HV at 350 °C. The AFM image reported in Fig. 2(a) clearly differs with respect to the pristine sample, even though SLG wrinkles are still clearly observable. Indeed, the height histogram (inset of Fig. 2(a)) has now the characteristic lognormal shape typical of rough films deposited onto a very flat substrate, and the *rms* roughness over the whole image now reads 1.2 nm.

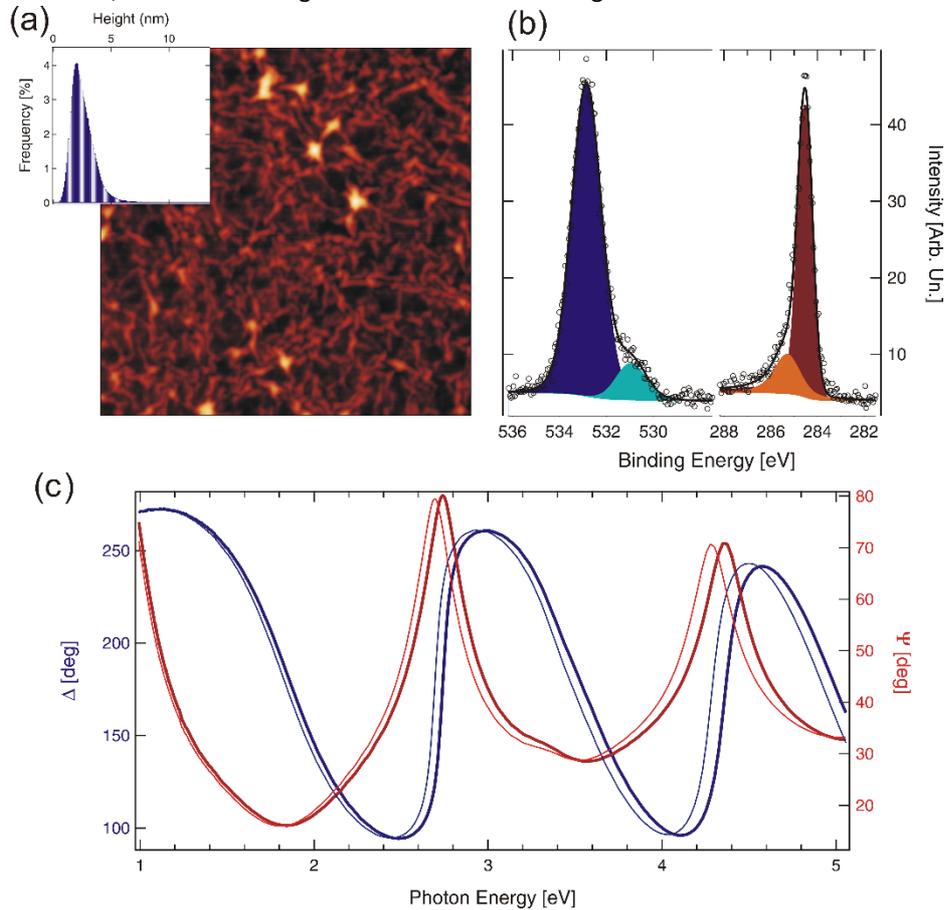

**Fig. 2.** Panel (a): AFM image of annealed SLG/SiO$_2$. Image size is 1x1 mm$^2$. Inset of panel (a): corresponding height-distribution histogram. Panel (b): high-resolution XPS spectra of the O1s (left) and C1s (right) spectral region of the annealed SLG/SiO$_2$ sample. The experimental data (markers), best fit (black line) and the spectral decomposition into subcomponents (coloured areas) are shown. The intensity scale was maintained identical to Fig. 1. Panel (c): ellipsometry spectra Ψ(E) (dark red line) and Δ(E) (deep-blue line) as a function of photon energy measured from the annealed SLG/SiO$_2$. The thin red (blue) line represents the pristine-SLG/SiO$_2$ data, and is reported to facilitate the observation of the annealing-induced modifications. The incidence angle was 66°.

The SE spectra are reported in Fig. 2(c) as the dark-red (Ψ) and deep-blue (Δ) lines. For the sake of comparison, the SE spectra of the pristine system are reported as the thin red and blue lines. A clear blueshift of the Ψ and Δ oscillations is observed after the annealing, with only very minor shape modifications. In practice, in terms of the *spectral position* of the Ψ, Δ oscillations, the annealed spectra revert to resembling the bare-substrate curves, indicating the decrease of the overall thickness of the dielectric layer located on top of the Si substrate.

The XPS spectra of the O1s and C1s core-level regions are reported in Fig. 2(b). The C1s graphitic subcomponent has remained constant in intensity, and became sharper, while the highest-BE subcomponent has disappeared. In the O1s case, the SiO2-related dominant subcomponent has increased in intensity and, out of the several subcomponents originally present, we can observe only a low-BE shoulder at 531 eV.

## 2.2.   Optical models of graphene

In order to quantitatively exploit SE, and conjure up a coherent interpretation for all the data recorded, it is necessary to devise optical models for the sample that allow to interpret the SE data. Such models typically consist of stacks of *N* layers, each characterized by its own thickness and dielectric function, with Fresnel boundary conditions for the electromagnetic field [12]. In this respect, one of the major issues encountered when dealing with SLG is that, due to its 2-dimensional character, a universally accepted representation of its dielectric function does not at present exist. Before describing our own approach, it is therefore instructive to briefly review the state of the art.

Among past SE investigations of SLG, Wang et al. (Ref. [19]) extracted the optical constants of graphene flakes using a patented dispersion formula based on Tauc-Lorentz model without, however, reporting the dispersion parameters [20]. Kravets et al. [21], instead, modelled graphene as an anisotropic material with arbitrary *xy* response and a Cauchy (*i.e.* transparent, dispersive) response for the *z* component of the dielectric tensor. The effective thickness of graphene flakes $t_{SLG}$ was fixed at its widely-accepted nominal value of $t_{SLG}$=0.335 nm, and a Cauchy layer was introduced as a spacer between graphene and the substrate, to account for the presence of water and/or air resulting from the transfer process. Just a few months later, Weber et al. proposed a model for graphene's dielectric function based on B-splines, fitting it to several experimental spectra simultaneously [20]. Running their fitting algorithm with several values for graphene thickness $t_{SLG}$, they obtained the best fit for $t_{SLG}$=0.34 nm. Nelson et al. proposed an isotropic model for polycrystalline CVD graphene based on Lorentzian oscillators only, showing little difference with respect to Ref. [21] and thus indicating that anisotropy plays little role in determining graphene's optical constants [22]. In 2011, Losurdo et al. demonstrated the real-time, in-situ SE analysis of graphene growth, extracting the pseudo-dielectric function of SLG [23], and comparing it to the available references [21,22]. In 2012, Matkovic et al., following SE measurements in the visible range (360-800 nm), calculated the optical constants of graphene using a phenomenological Fano model [24] and reported the presence of a thin ($\approx$1 nm) Cauchy layer *on top* of SLG [25]. Chang et al. [26] shifted the paradigm and modelled this material as an infinitely thin layer with an in-plane optical conductivity, thereby expressing their results in terms of optical conductivity and claiming that in a truly 2D material, the refractive index is an ill-defined quantity [26]. A more conservative approach was then again presented by Losurdo et al., aiming to explain the usefulness of SE applied to graphene characterization [27]; there, the optical constants of graphene were extracted either via a point-by-point fit, or a Drude-Lorentz model, indicating an optimal effective thickness $t_{SLG}$ value of 0.34 nm, the same already observed by Weber [20]. Beside the adoption of so many different approaches, matters are further complicated by the fact that different SLG fabrication techniques lead to different results in terms of SLG quality and morphology, thereby influencing the effective properties of the system [13]. For example, no intercalation layer is reasonably expected in exfoliated or dry-transferred SLG, whereas it might be present in wet-transferred samples.

With these prerequisites, we have modelled our systems as follows (see Fig. 3(a)): bottom to top, we introduced the Si substrate, a $SiO_2$ film, a Cauchy layer (henceforth also referred to as "intercalation layer"), then SLG with its own surface-roughness layer (not represented in the figure). For Si and $SiO_2$, dielectric functions from the literature were employed [28]. The thickness of the Si-oxide was determined via best fits of the SiO2/Si system *prior to* the transfer of SLG, and later maintained fixed at the best-fit value of 290.5 nm. For the intercalation layer, we assumed the dielectric function of water, modelled via the Cauchy approximation, since our samples were wet-transferred in acqueous solution. A so-called intermixing layer was introduced between the $SiO_2$ and the intercalation layer in order to account for the wafer's surface roughness; its thickness was fixed at 2.7 nm based on the best-fit analysis of the $SiO_2$/Si substrate prior to the transfer of SLG.

Graphene was modelled as the sum of two functions: a Lorentz oscillator and a parametrized function called PSEMI [29]. The PSEMI combines a highly flexible functional shape with Kramers-Kronig consistency. Its flexibility has promoted its use in contexts where complex dielectric peak shapes have to be reproduced [30,31]. It is therefore particularly suited to reproduce the asymmetric peak that arises around 4.6 eV in the optical conductivity of graphene; the reader interested in the details of the PSEMI function is referred to the WVASE® User's Guide by J. Woollam Co., Inc. [32]. The Lorentz function has been centered in the near-IR to describe the vertical interband transitions between the valence and conduction band of graphene; the PSEMI, centered in the UV, accounts for the interband transition peak [33,34]. The thickness of graphene was fixed at 0.335 nm.

The fitting procedure involved first the adaptation of the PSEMI functional to reproduce the graphene's dielectric response on reference samples, then the actual fitting of the samples presented in this work. The first step was performed by fitting the SE response of SLG samples where AFM and XPS data allowed to exclude the presence of intercalation layers. All the parameters of the Lorentz and PSEMI functions were fitted except for the PSEMI central energy, which was fixed at the literature value of 4.6 eV [35]. The results obtained were systematically tested against literature data, yielding good agreement. The parameters thus obtained served as the starting point for the fitting of the data presented here, for which independent indications of non-idealities such as the intercalation layer were obtained. For this, the thickness of the SLG surface roughness and of the intercalation layer were left free to vary, while only fine adjustments were allowed for the Lorentz and PSEMI functions, meant to account for small sample-to-sample variations in the dielectric response of SLG. The real and imaginary part of the refractive index of SLG employed in this work are reported in Fig. 3(b).

The best fit procedure for the pristine SLG yielded a 5.2 nm-thick intercalation layer, a 0.4 nm-thick roughness layer for SLG and was obtained assuming the intermixing layer to be composed of equal parts of $SiO_2$ and intercalant. The corresponding experimental SE data (markers) and the best fit (lines) are reported in Fig. 4, top. For the annealed sample, the best fit yielded an unchanged SLG roughness, no intercalation layer and the consequent replacement of the intercalant fraction in the intermixing layer with voids (*i.e.* a material with *ε*=1). The corresponding SE data (markers) and best fit (lines) are reported in Fig. 4, bottom.

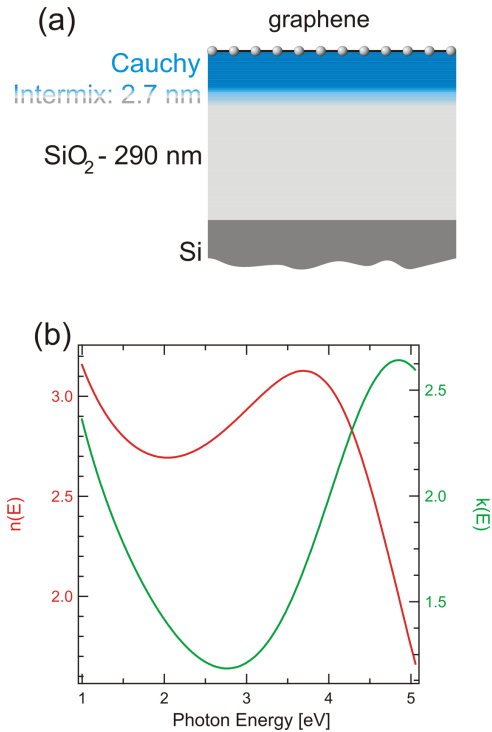

**Fig. 3.** Panel (a): sketch of the optical model employed to fit the ellipsometry data. Panel (b): dielectric functions of graphene employed in the present paper.

## 2.3. Discussion

The combined AFM, XPS and SE evidence concurs in providing a coherent model of the SLG/SiO$_2$ system both before and after the annealing. In the pristine case, AFM indicates that the outermost sample surface lies on two different height levels, separated in height by 2.5 nm. The small roughness of the lowest level suggests that the bottom level is composed of SLG directly adhered to the SiO$_2$ surface, whereas an intercalation layer is present between SLG and the substrate in the form of a series of nanopockets with limited lateral extension and well-defined height, bounded by the SLG areas adhered to the substrate. Assuming that the bottom level corresponds to SLG laid directly on the substrate, the averaged intercalation-layer thickness deduced by AFM is slightly below 2 nm. The XPS spectrum in the O1s region is dominated by the SiO$_2$ subcomponent (dark blue, 532.9 eV BE), with a high energy shoulder assigned to H$_2$O (fucsia, 533.4 eV) and more peaks possibly related to minor residual graphene contamination (confirmed by the presence of non-graphitic subcomponents in the C1s spectrum) [36,37] (light blue). The H$_2$O-related subcomponent points to water as the most likely intercalant material [38], as indeed conceivable considering that the transfer was performed in acqueous solution; its relatively weak signal with respect to the SiO$_2$ peak reinforces the hypothesis that the bottom layer in the AFM images corresponds indeed to SLG laid directly on the substrate, since a thicker intercalation layer would have more-strongly damped the XPS signal from the substrate. In correspondence of this, the SE model indicates the presence of an intercalation layer (5.2 nm of H$_2$O) between SLG and the substrate.

Following the annealing, all the data testify the occurrence of significant modifications in the system [5]. AFM indicates that the outer surface now corresponds to a morphological single-level system, with little residual roughness, and that the characteristic nanopockets have disappeared. In correspondence of this, XPS shows a significant increase of the SiO$_2$-related signal in the O1s spectrum, accompanied by the loss of the H$_2$O component and a variation of the adventitious signals. The graphitic C1s peak slightly sharpens, possibly due to the annealing of defects, and remains unchanged in intensity, while the highest-BE adventitious component disappears [7,37]. Correspondingly, the SE model achieves the best fit with the experimental data assuming that the intercalation layer thickness has dropped to zero.

Overall, the data are readily interpreted assuming the annealing-induced evaporation of the intercalation layer and the desorption of most adventitious molecules [7]. Such an evaporation clearly leads to the loss of the nanopockets, hence the change in the AFM image. Additionally, it justifies the disappearance of the $H_2O$ peak in XPS and the blueshift of the Ψ, Δ peaks. The fact that the graphitic C1s peak remains unchanged in intensity after the annealing while the $SiO_2$ component increases demonstrates that the evaporation of material occurs from *below* the SLG layer, yet above the $SiO_2$, otherwise both peaks should have shown analogous behaviour. Thus, all the different data sets provide a fully coherent picture of the system and, most importantly, SE nicely fits in this frame and clearly shows a predictive character in terms of the identification of an intercalation layer. Our results are consistent with recent data showing that vacuum annealing has a relevant influence in modifying the degree of adhesion of SLG to target substrates after wet-transfer procedures [39], with important fallouts for applications. We just notice the difference recorded between the thickness of the intercalation layer extracted by SE and AFM (5 nm *vs* 2 nm). Such a discrepancy may have different origins, like the difference in area sampled by AFM or SE, or the uncertainty level inherent to the SE modelling of graphene-based systems, or finally a confinement effect on the intercalant dielectric constant. Ruling out the latter option, since the typical pressures in graphene bubbles is not extreme [40], it is possible that the dielectric peculiarities of SLG-based systems might somewhat affect the precision of SE modelling. The central point is however that SE data appear conclusive and predictive in correctly determining the presence (and removal) of intercalation layers trapped between the substrate and graphene. Once this principle is established, SE may be applied in a stand-alone configuration for assessing the morphological quality of this class of systems.

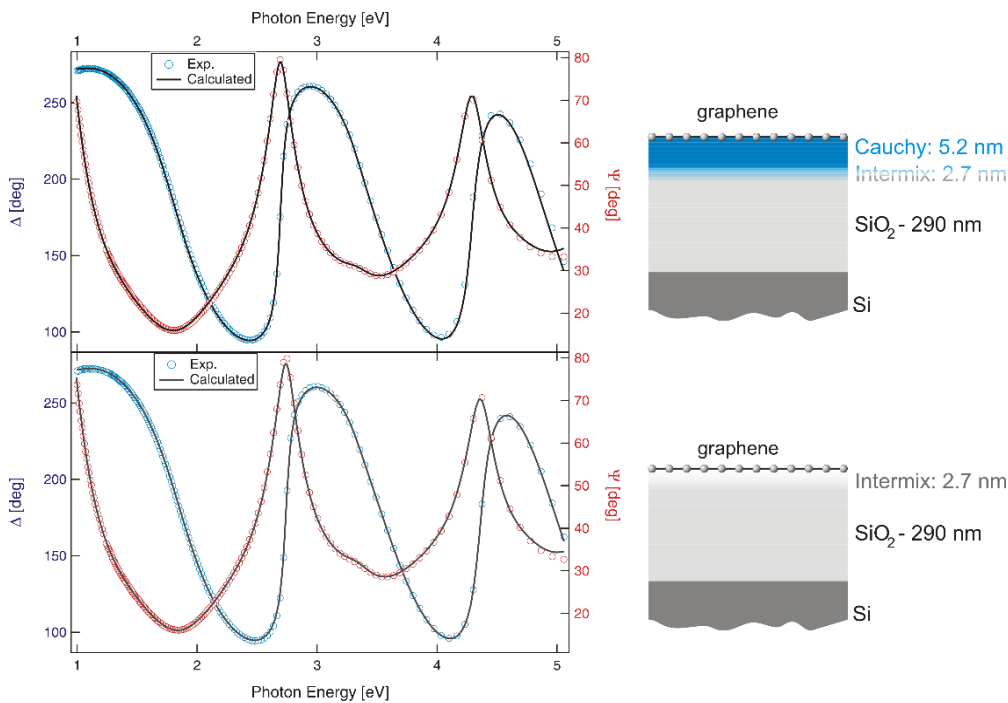

**Fig. 4**. Top: pristine SLG/SiO$_2$, experimental (markers) and calculated (lines) ellipsometry spectra. The corresponding model is reported on the right side. Bottom: annealed SLG/SiO$_2$, experimental (markers) and calculated (lines) ellipsometry spectra. The corresponding model is reported on the right side. The experimental Ψ(*E*) and Δ(*E*) spectra are reported as red and blue markers, respectively.

## 3. Conclusion

Summarizing, we have performed a multitechnique investigation of the interface morphology of large-area polycrystalline graphene transferred on a SiO$_2$/Si wafer by wet methods. We have exploited spectroscopic ellipsometry, backed by AFM and XPS, for investigating the presence of residual spurious intercalation layers between the substrate and SLG. On a selected number of samples showing a peculiar two-level surface morphology, we found that in order to achieve a good fit of the spectroscopic ellipsometry response of freshly-transferred SLG it was necessary to introduce an optically-transparent intercalation layer between SLG and the substrate. Following a mild annealing of the system in high vacuum, we observed the disappearance of the two-level surface morphology, and correspondingly that the ellipsometry response of the whole system could now be reproduced without resorting to intercalation layers. Ancillary XPS data allowed to ascribe the intercalation layer to residual water from the wet-transfer procedure, and confirmed that such a water layer was indeed trapped between the substrate and the SLG. This study shows the capability of SE to correctly capture the presence of few-nm thick spurious intercalated layers between graphene and substrate following a wet-transfer procedure. Given the typical speed of acquisition of SE spectra (few seconds), the non-invasive and non-destructive character of the technique and its applicability in air, vacuum and liquid environments, our study promotes ellipsometry as a fast and non destructive method for evaluating the quality of SLG adhesion following wet-transfer processes, either for *ex-post* or in-line quality evaluation.

**Acknowledgements**

Financial support from the Ministero dell'Istruzione, Universita e Ricerca (Project no. PRIN 20105ZZTSE_003) is acknowledged. Part of the research leading to these results has received funding from the European Union Seventh Framework Program under grant agreement no. 604391 Graphene Flagship. The author thank Prof. Alberto Morgante for stimulating discussions. The image analysis was performed using the open-source software Gwyddion [41].